\documentclass{appolb}
\usepackage{epsfig}

\begin{document}
\title{Transport Coefficients and $n$PI methods.%
\thanks{Talk presented at the HIC for FAIR Workshop and
XXVIII Max Born Symposium ``Three Days on Quarkyonic Island,'' Wroclaw, May 19-21, 2011.}
%
}
\author{M.E. Carrington 
\address{Department of Physics, Brandon University, Brandon, Manitoba, R7A 6A9 Canada\\ and \\
  Winnipeg Institute for Theoretical Physics, Winnipeg, Manitoba}
}
\maketitle
\begin{abstract}
Transport coefficients can be obtained from 2-point correlators using the Kubo formulae. It has been shown that the full leading order result for electrical conductivity and (QCD) shear viscosity is contained in the re-summed 2-point function that is obtained from the 3-loop 3PI  effective action. The theory produces all leading order contributions without the necessity for power counting, and in this sense it provides a natural framework for the calculation and suggests that one can calculate the next-to-leading contribution to transport coefficients from the 4-loop 4PI effective action. The integral equations have been derived for shear viscosity for a scalar theory with cubic and quartic interactions, with a non-vanishing field expectation value. We review these results, and explain how the calculation could be done at higher orders. 
\end{abstract}
  
\section{Leading order transport coefficients from the $n$PI effective action}

It is well known that selective resummations can play an important role in quantum field theory. A prominent example is the hard thermal loop theory, which includes screening effects that regulate infra-red divergences. The convergence of such perturbation theories can typically be improved using an $n$-particle irreducible ($n$PI) effective action. In addition, $n$PI approximation schemes can be used to study far from equilibrium systems. 


Although the $n$PI effective action is consistent with the global symmetries of the theory, symmetry identities for $n$-point functions may not be satisfied during intermediate steps of the  calculation \cite{AS,HZ}. For scalar theories, Goldstone's theorem may not be satisfied, and for gauge theories, Ward identities can be violated. To address this problem we look at the resummed $n$PI effective action, which is defined with respect to the self consistent solutions of the $n$-point functions. This type of strategy was originally proposed by Baym and Kadanoff \cite{baym}. 
The resummed effective action respects all symmetry properties of the theory, and the $n$-point functions which are obtained by functional differentiation satisfy Goldstone's theorem, or the Ward identities. 

Transport coefficients can be calculated from the Kubo formulae. Conductivity and shear viscosity are obtained from the zero frequency limit of the 2-point function. This limit produces pinch singularities, which come from pairs of retarded and advanced propagators which carry the same momenta. When integrating a term of the form $\int dp_0 \;G^{ret}(P)G^{adv}(P)$, the integration contour is `pinched' between poles on each side of the real axis, and the integral contains a divergence called a `pinch singularity.' These divergences are regulated by using resummed propagators which account for the finite width of thermal excitations. This procedure introduces extra factors of the coupling in the denominators which change the power counting. As a consequence, in even the leading order (LO) calculation, there is an infinite set of graphs which contain products of pinching pairs that need to be resummed. 
This is accomplished by solving an integral equation whose kernel is the square of the matrix elements that correspond to the 2 $\rightarrow$ 2 scattering and production processes. In gauge theories, in addition to pinch singularities, the presence of collinear singularities makes 2 $\rightarrow$ 3 scatterings as important as 2 $\rightarrow$ 2 scatterings. These collinear terms are resummed by another integral equation. 
The complete LO result is obtained by solving the two coupled integral equations.

The integral equations that resum pinch and collinear singularities can be obtained from the equations of motion (eom's) of the  3-loop 3PI effective action \cite{EK1,EK2,EK3}.
Gauge invariance is automatically satisfied, and all LO terms appear without the need for any kind of power counting arguments. 

\section{Next to leading order transport coefficients}

Little progress has been made on the calculation of transport coefficients beyond LO (see however \cite{moorePhi4,simon1,simon2}).
Since power counting is notoriously difficult, and becomes increasingly complicated at higher orders, $n$PI effective theories could provide a useful method to organise the calculation of transport coefficients at next-to-leading order (NLO).

 In spite of the complexity of the $n$PI effective action, a systematic expansion can be done in a self-consistent way \cite{berges}. 
Using the 4-loop 4PI effective action for a scalar theory with cubic and quartic interactions, with non-vanishing field expectation value, one can derive the integral equations that would produce the NLO order contribution to the viscosity \cite{EK4}. We remark that the numerical solution of these equations is more difficult than anything that has been accomplished so far. In addition,  renormalizability has not been demonstrated beyond the level of the 2PI effective action \cite{reinosaRenorm1,reinosaRenorm2}.
 
\begin{figure}[H]
\begin{center}
\includegraphics[width=15cm]{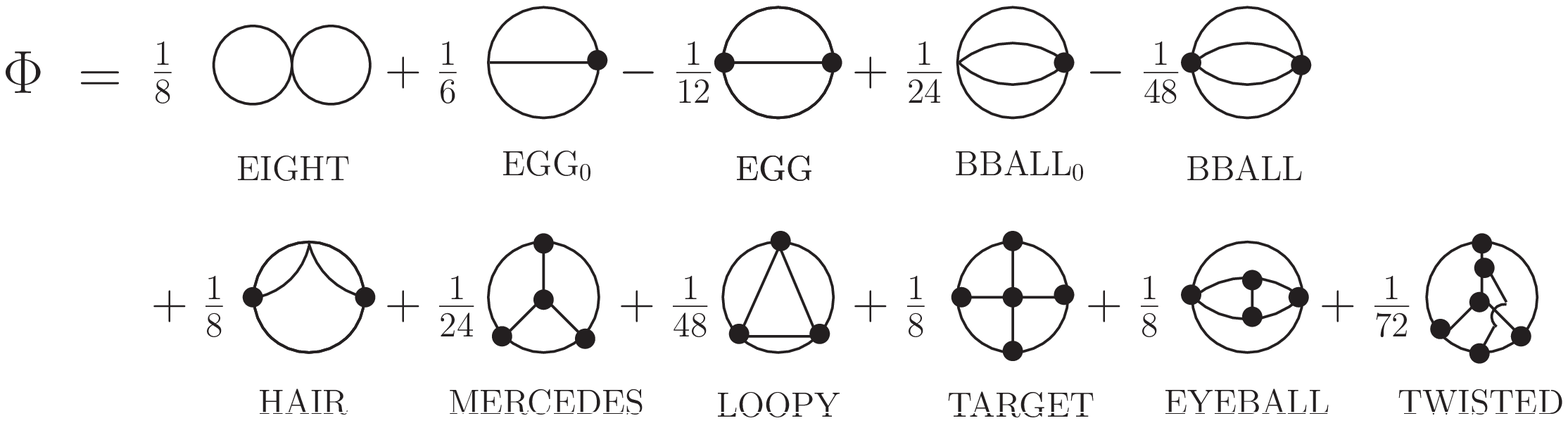}
\end{center}
\caption{\label{PHI}4-loop 4PI effective action.}
\end{figure}

We use a compactified notation in which a single numerical subscript represents all space-time co-ordinates. For example: the field expectation value is written $\phi_1:=\phi(x)$, 
the variational propagator is written $D_{12}:=D(x_1,x_2)$, the bare 4-point vertex is written $V^0_{1234}:=V^0(x_1,x_3,x_2,x_4)$, etc.  We also use an Einstein convention in which a repeated index implies an integration over space-time variables. 
Using this notation we write the classical action:
\begin{eqnarray}
\label{scl}
S_{cl}[\phi]=\frac{1}{2}\phi_1\big[i\,(D^{0}_{12})^{-1}\big]\phi_2-\frac{i}{\;3!} U_{123}^{oo}\phi_1\phi_2\phi_3-\frac{i}{\;4!}V_{1234}^0\phi_1\phi_2\phi_3\phi_4\,.
\end{eqnarray}
The 4PI effective action has the form:
\begin{eqnarray}
\label{Gamma4PI}
&&\Gamma[\phi,D,U,V]
=S_{cl}[\phi]+
    \frac{i}{2} {\rm Tr} \,{\rm Ln}D^{-1}_{12}\\
    && +
\frac{i}{2} {\rm Tr}\left[(D^0_{12}(\phi))^{-1}\left(D_{21}-D^0_{21}(\phi)\right)\right]-i\Phi^0[\phi,D,U,V]-i\Phi^{\rm int}[D,U,V] \,. \nonumber
\end{eqnarray}
The terms $\Phi^0[\phi,D,U,V]$ and $\Phi^{\rm int}[D,U,V]$ contain all contributions to the effective action which have two or more loops.  The first piece $\Phi^0[\phi,D,U,V]$ includes all terms that contain  bare vertices. 
The diagrams up to 4-loops are shown in Fig. \ref{PHI}.
\par
There are four eom's which are obtained by functionally differentiating with respect to the four functional arguments of the effective action:
\begin{eqnarray}
\label{eom1}
&&\frac{\delta \Gamma[\phi,D,U,V]}{\delta X_i}=0\,,~~~~X_i \in \{\phi,D,U,V\}\,.
\end{eqnarray}
The equations obtained by varying with respect to $\{D,U,V\}$ can be solved simultaneously for the self-consistent solutions which are functions of the field expectation value: $\tilde D[\phi]$, $\tilde U[\phi]$, $\tilde V[\phi]$.
Substituting these self-consistent solutions we 
obtain the resummed action, which depends only on the expectation value of the field:
\begin{eqnarray}
\label{Gamma4PI-rs}
\tilde{\Gamma}[\phi]=
\Gamma[\phi, \tilde{D}[\phi], \tilde{U}[\phi], \tilde{V}[\phi]]
\,.
\end{eqnarray}

\begin{figure}
\begin{center}
\includegraphics[width=16cm]{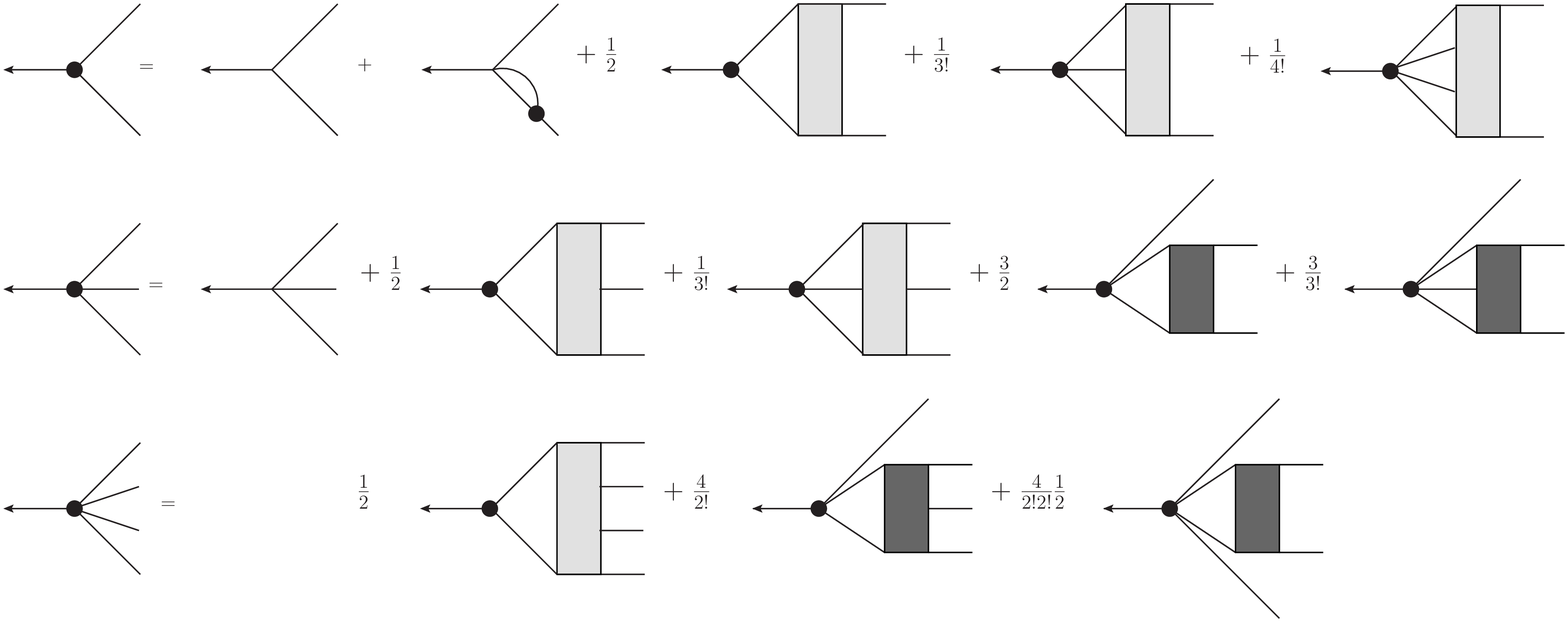}
\end{center}
\caption{\label{OMPSint}The integral equations for the vertices $\Omega$, $\Psi$ and $\Theta$ defined in equation (\ref{vert-defns}). The kernels are drawn as grey boxes.}
\end{figure}

\begin{figure}
\begin{center}
\includegraphics[width=12cm]{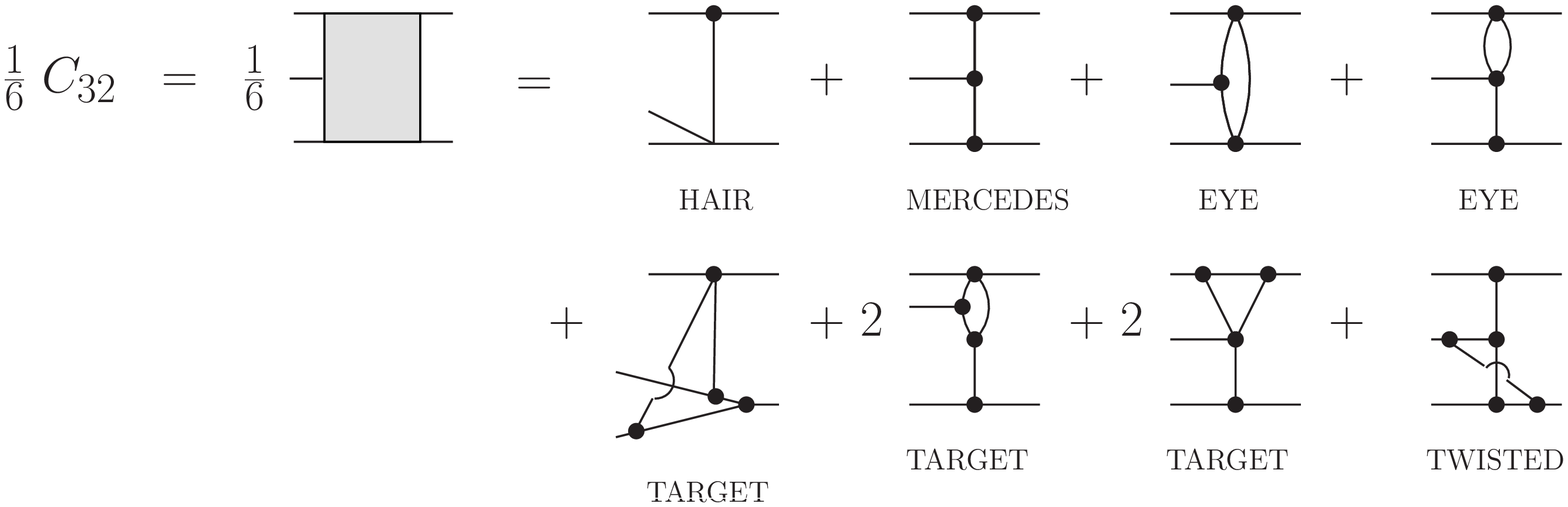}
\end{center}
\caption{\label{Ceqn}One of the kernels in Fig. \ref{OMPSint}.}
\end{figure}

\begin{figure}
\begin{center}
\includegraphics[width=16cm]{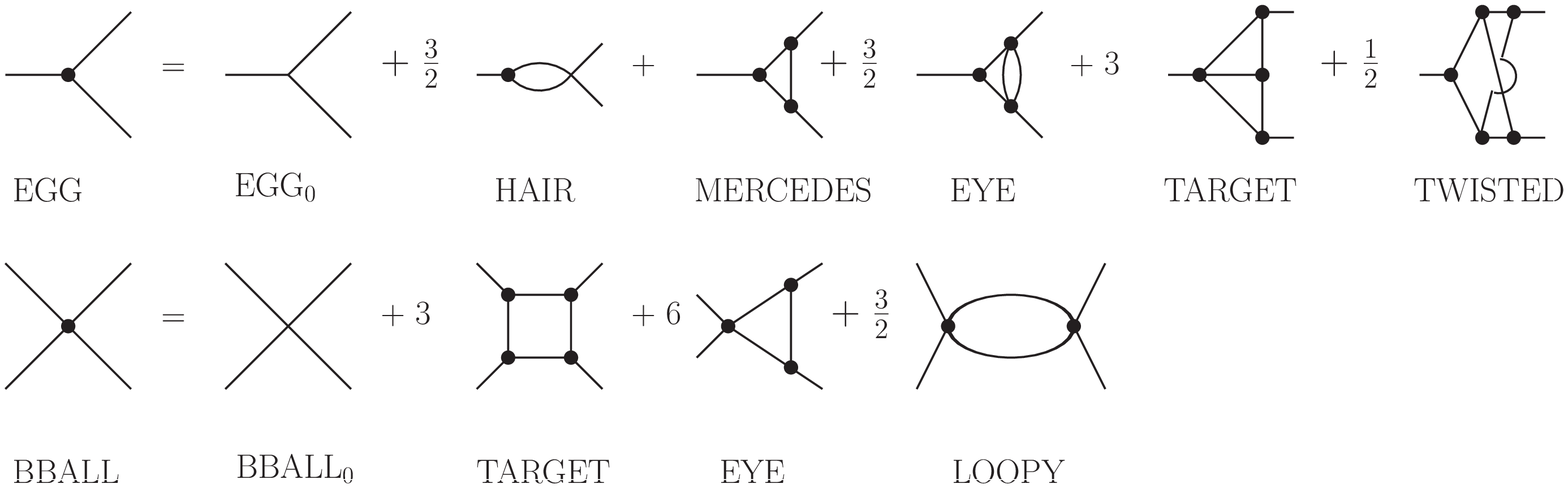}
\end{center}
\caption{\label{Uint}Integral equations for the variational vertices $U$ and $V$ from the 4-Loop 4PI effective action.}
\end{figure}

The re-summed propagator and self energy are defined as: 
\begin{eqnarray}
\label{ext-prop}
i(D^{{\rm ext}}_{12})^{-1}=   
     \frac{\delta^2}{\delta \phi_2 \delta \phi_1}
     \tilde{\Gamma}[\phi]=i\big[(D^{0}_{12}(\phi))^{-1}- \Pi^{\rm ext}_{12}\big]\,.
\end{eqnarray}
We can derive an expression for the re-summed 2-point function by taking derivatives of the resummed effective action and using the chain rule. The result has the same form as the Schwinger-Dyson (sd) equation and depends on the 3- and 4-point vertices defined in (\ref{vert-defns}). We also define a 5-point vertex that we will need below.
\begin{eqnarray}
\label{vert-defns}
&&\Omega_{123} = -\frac{\delta \tilde{D}^{-1}_{12}}{\delta \phi_{3}}\,,~~~~\Psi_{1234}=\frac{\delta \tilde U_{123}}{\delta\phi_4}\,,~~~~
\Theta_{12345}=\frac{\delta \tilde V_{1234}}{\delta\phi_5}\,.
\end{eqnarray}
The vertices in equation (\ref{vert-defns}) satisfy integral equations obtained by taking functional derivatives with respect to the field expectation value of the appropriate eom (\ref{eom1}). These integral equations are shown in Fig. \ref{OMPSint}. One of the kernels is shown in Fig. \ref{Ceqn}. The integral equations for the vertices in (\ref{vert-defns}) depend on the vertices $U$ and $V$, which in turn satisfy their own integral equations, that are obtained directly from (\ref{eom1}) and shown in Fig. \ref{Uint}. In all figures we combine diagrams that correspond to permutations of external legs. The full NLO contribution to the shear viscosity can be obtained from the Kubo formula where the 2-point function is given by the sd equation with vertices obtained by solving the set of coupled integral equations shown in Figs. \ref{OMPSint} and \ref{Uint}.
\par
\par
\par

\section{A new approach to the calculation of the effective action}
\label{newApproachSection}

It is  interesting to consider how the calculation would proceed at higher orders. 
The $n$PI effective action is defined as the $n$th Legendre transform of the generating functional obtained by coupling to $n$ source terms. 
A direct calculation beyond the 4-loop 4PI level is extremely difficult \cite{Yun5}, but it is possible to calculate the effective action without taking any Legendre transforms \cite{Yun6}.

The key to the method is the introduction of a set of fictitious bare vertices: to obtain the $n$-Loop $n$PI effective action we include in the Lagrangian the vertices $V_j^{0}$ for $j=3,4,5,6,\cdots\,n$. The inclusion of the non-renormalizable interactions ($j\ge 5$) is an organizational trick, and these vertices are set to zero at the end of the calculation. From this point on, we consider $\phi=0$ for simplicity, and use the notation $V_j^0$ and $V_j$ to represent bare and proper vertices with $j$ legs. 
Using these fictitious vertices, we can show that the eom's and sd equations are equivalent to the order at which the truncated theory respects the symmetries of the original theory.\footnote{The sd equations and eom's are equivalent only when fictitious vertices are used, and this result is important {\it only} because it allow us to obtain the effective action without taking a series of Legendre transforms. It is {\it not} true that the non-perturbative solutions of a truncated set of sd equations are the same as the solutions of the eom's obtained from the $n$PI effective action. 
}
This result allows us to construct the $n$-Loop $n$PI effective action directly from the sd equations. The method has been used to reproduce, with comparatively little effort, the known results for the $n$-Loop $n$PI effective action with $n=4$ and $n=5$. The technique has also been used to calculate the 6-Loop 6PI effective action which is, realistically speaking, impossibly tedious to obtain using Legendre transforms.

The first step is to compare the perturbative expansions of the sd equations and the eom's. The complete set of eom's can be written:
\begin{eqnarray}
\label{EOM2}
&& V_j = V_j^0 + {\rm fcn}^\prime_j[V_l^0,V_k]+ {\rm fcn}_j[V_k]\,.
\end{eqnarray}
The functions ${\rm fcn}^\prime_j[V_l^0,V_k]$ and ${\rm fcn}_j[V_k]$ are defined as:
\begin{eqnarray}
\label{EOMforce}
j=2:~~&&{\rm fcn}^\prime_2[V_l^0,V_k] =  - 2 \frac{\delta \Phi^0[V_l^0,V_k]}{\delta D}\,,~~~~{\rm fcn}_2[V_k] =  - 2\frac{\delta  \Phi^{\rm int}[V_k]}{\delta D}\,,\\
j\ge 3:~~&&{\rm fcn}^\prime_j[V_l^0,V_k] =   j! D^{-j}\frac{\delta \hat\Phi^0[V_l^0,V_k]}{\delta V_j}\,,~~~~{\rm fcn}_j[V_k] =   j! D^{-j}\frac{\delta  \hat\Phi^{\rm int}[V_k]}{\delta V_j}\,,\nonumber
\end{eqnarray}
where the hats indicate that the `basketball' diagrams (for example the 2nd-5th diagrams in Fig. \ref{PHI}) which produce the tree terms in (\ref{EOM2}) are dropped. 
The sign difference for the 2-point function and the missing factor $D^{-2}$ occurs because of the fact that it is conventional to write the effective action as a function of the propagator instead of the inverse propagator.

 We can generate the perturbative expansion of any functional of proper vertices by repeatedly substituting (\ref{EOM2}). We can also repackage a perturbative set of diagrams as skeleton diagrams by repeatedly using the same equation in the form:
\begin{eqnarray}
\label{subbereom}
&&V^0_j= V_j-{\rm fcn}^\prime_j[V_l^0,V_k]-{\rm fcn}_j[V_k]\,.
\end{eqnarray}

\begin{description}
\item[item 1] If we convert a set of skeleton diagram for the vertex $V_j$ into a series of perturbative diagrams using (\ref{EOM2}), the leading loop order of the new set of diagrams is greater than or equal to the leading loop order of the original set.
\item[item 2] If we include fictitious vertices $V^0_j$ for $5\le j \le n$, we can convert skeleton diagrams to perturbative diagrams using (\ref{EOM2}), or perturbative diagrams to skeleton diagrams using (\ref{subbereom}), and the leading loop order of the new set of diagrams is equal to the leading loop order of the original set.
\end{description}

\begin{figure}
\begin{center}
\includegraphics[width=11cm]{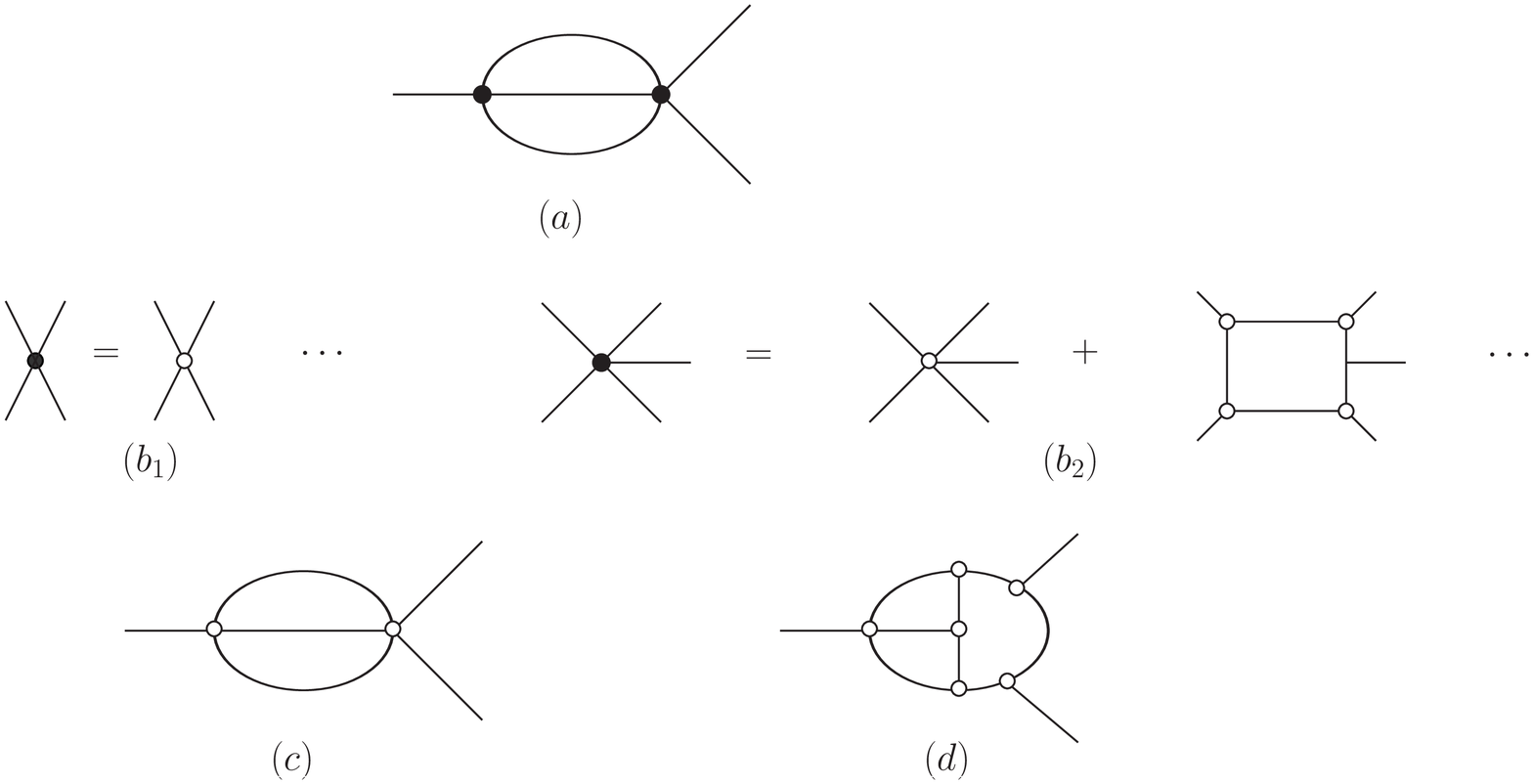}
\end{center}
\caption{\label{ptExample} Diagrams used to explain items 1 and 2 in section \ref{newApproachSection}.}
\end{figure}

We illustrate these statements with an example. We use $L_{pt}$ to indicate the loop order of the perturbative expansion.
Consider the skeleton diagram shown in part $(a)$ of Fig. \ref{ptExample}, which is of order $L=2$. We can expand this diagram as a series of perturbative diagrams using equations of the form (\ref{EOM2})  which are shown for this example in part $(b)$ of the figure.\footnote{The propagators in the skeleton diagrams in Figure \ref{ptExample} also have to be expanded to obtain a perturbative diagram. This will produce extra loops that correspond to self energy corrections. In this paper we do not introduce notation to distinguish skeleton and perturbative propagators in diagrams.} The LO term is shown in part $(c)$, and is of order $L_{pt}=2$. Thus we have $L=L_{pt}=2$. Now consider the result if we set $V_5^0=0$, which means we remove the first diagram on the right side of part $(b_2)$. In this case the LO term is shown in part $(d)$ and is of order $L_{pt}=3$. Thus we see that if the fictitious vertex $V_5^0$ is set to zero we have $L_{pt} > L$.
\par

We consider truncating the $n$PI effective action at $m$-loop order.\footnote{At $m$-Loops, the $n$PI effective action is the same as the $(n+1)$PI effective action for $n>m+1$ \cite{berges}.} The functional derivative of an $m$-loop graph with respect to the variational vertex $V_j$ opens $j-1$ loops. This means that an arbitrary $m$-loop graph in the effective action which contains the vertex $V_j$ produces a term with ${\cal L}[m,j]$ loops in the skeleton expansions of the eom for the vertex $V_j$, where we  define:
\begin{eqnarray}
\label{calL}
{\cal L}[m,j] := m-j+1\,.
\end{eqnarray}

Now we consider the Schwinger-Dyson equations, which form an infinite hierarchy of coupled non-linear integral equations. They have the form:
\begin{eqnarray}
\label{sdGeneral}
V_j^{sd} = V_j^0+{\rm fcn}_j^{sd}[V_l^{0},V_k^{sd}]\,.
\end{eqnarray}
Although the structure of the sd equations is very different from the eom's, it is possible to show \cite{Yun5}:
\begin{description}
\item[item 3] The perturbative expansions of the $n$-loop $n$PI eom's and the sd equations truncated by setting $V^{sd}_{n+k}=V^{0}_{n+k}$ for $k\ge 1$ both match the perturbative expansion obtained from the 1PI effective action, and therefore each other, to order $L_{pt}= {\cal L}[n,j]$.
\end{description}
This result is perhaps not surprising, since both formalisms should contain the correct perturbative physics up to the order at which the calculation is done. Item 3 tells us that both  the variational and sd vertex functions respect crossing symmetry to order $L_{pt}= {\cal L}[n,j]$, which we refer to as the ``truncation order.''

We can formally write equation (\ref{sdGeneral}) as:
\begin{eqnarray}
\label{newsd}
&& V_j^{sd} = V_j^0+{\rm fcn}^{\prime}_j[V_l^0,V_k^{sd}]+ I_j[V_l^0,V_k^{sd}]\,,\\[2mm]
&& I_j[V_l^0,V_k^{sd}]:= {\rm fcn}_j^{sd}[V_l^0,V_k^{sd}]-{\rm fcn}^{\prime}_j[V_l^0,V_k^{sd}]\,.\nonumber
\end{eqnarray}
Comparing (\ref{EOM2}) and (\ref{newsd}) and using item 3, it is clear that ${\rm fcn}_j[V_k]$ and $I_j[V_l^0,V_k^{sd}]$ must match each other in the perturbative expansion to order $L_{pt}={\cal L}[n,j]$.
Therefore we can rewrite $I_j[V_l^0,V_k^{sd}]$ as:
\begin{eqnarray}
\label{sub}
I_j[V_l^0,V_k^{sd}]={\rm fcn}_j[V_k^{sd}]~+ ~{\rm extra}\,,
\end{eqnarray}
where the extra term is of order $L_{pt}={\cal L}[n,j]+1$.
Using item 2 \emph{which is only true in the presence of fictitious bare vertices}, the extra term can be rewritten as a series of skeleton diagrams of order $L={\cal L}[n,j]+1$. Thus we have shown that:
\begin{description}
\item[item 4] \emph{In the presence of fictitious bare vertices} the  sd equations can be rearranged to have the same form as the $n$PI eom's, plus additional terms of order $L= {\cal L}[n,j]+1$ in the skeleton expansion, which is beyond the truncation order.
\end{description}

It is straightforward to use the result in item 4 to calculate the $n$-Loop $n$PI effective action from the sd equations. 
First use (\ref{subbereom}) to remove bare vertices in $I_j$ to order ${\cal L}[n,j]$ for $n\ge j \ge 3$. 
Then join the legs of the resulting equation for $I_j$, set $V_l=0$ for $n\ge l \ge j+1$, and multiply each diagram by the symmetry factor:
\begin{eqnarray}
\label{calS}
S[j]=s(1/v_j)(1/j!)\,,
\end{eqnarray}
where $s$ is the numerical factor in front of the diagram in the $I_j$, and $v_j-1$ is the number of times the vertex $V_j$ appears in the diagram.

It is clear that joining the legs in diagrams in the eom's will produce all graphs in the effective action at a given order. The trick is to obtain the correct symmetry factor. Consider starting from a known result for the effective action $\Phi^{\rm int}$, taking derivatives of each graph with respect to each variational vertex, and trying to reconstruct the effective action by joining the legs in each of these eom's. 
A given graph in the effective action will produce contributions to the eom's of each vertex it contains. In order to produce the correct symmetry factor when joining legs, we must drop the corresponding contribution in all but one eom, which we can take to be the eom for the largest vertex present. This is accomplished by imposing the condition $V_l=0$ for $n\ge l \ge j+1$. If the largest vertex appears in a given diagram in the effective action more than once, the graph that is produced by joining the legs will have a symmetry factor that is too large by a factor equal to the number of times the vertex appears. The correct symmetry factor is produced by including the factor $(1/v_j)$ in equation (\ref{calS}). 

Following the simple procedure described above, one can reproduce the known results for the 4-Loop 4PI and 5-Loop 5PI effective actions. The method has also been used to calculate the 6-Loop 6PI effective action \cite{Yun6}.

\section{Conclusions}

The LO calculation of transport coefficients involves infinite resummations of diagrams that can be identified by power counting arguments. The integral equations that resum these diagrams are produced directly from the equations of motion of the 3-loop 3PI effective action. This result leads to the conclusion that $n$PI effective theories provide a natural framework for the calculation of transport coefficients beyond leading order. The calculation of next-to-leading order shear viscosity in scalar theory can be formulated using the 4-loop 4PI effective action. Higher order effective actions can be calculated in a efficient way by working with the Schwinger-Dyson equations. 

\end{document}